\pdfoutput=1

\documentclass[11pt]{article}

\usepackage{acl}

\usepackage{times}
\usepackage{latexsym}
\usepackage{arydshln}
\usepackage{graphicx}
\usepackage{subcaption}
\usepackage[T1]{fontenc}

\usepackage[utf8]{inputenc}
\usepackage{xcolor}

\usepackage{microtype}

\usepackage{inconsolata}

\usepackage{graphicx}

\usepackage{microtype}
\usepackage{url}
\usepackage{booktabs}
\usepackage{times}
\usepackage{arydshln}
\usepackage{subcaption}
\expandafter\def\csname ver@subfig.sty\endcsname{}
\usepackage{svg}
\usepackage{latexsym}
\usepackage[utf8]{inputenc}
\usepackage{tcolorbox}
\tcbuselibrary{skins,breakable}
\usepackage{inconsolata}
\usepackage{verbatim}
\usepackage{xspace}
\usepackage{subfig}
\usepackage{blindtext}
\usepackage{algpseudocode}
\usepackage{amsmath,amssymb,amsfonts}
\usepackage{cite}
\usepackage{amsmath,amssymb,amsfonts}
\usepackage{algorithm}
\usepackage{booktabs} 
\usepackage{siunitx} 
\usepackage{amsmath} 
\usepackage{graphicx} 
\usepackage{float} 
\usepackage{xcolor}
\usepackage{multirow}
\usepackage{algorithm,algpseudocode}
\usepackage{listings}
\definecolor{codegray}{rgb}{0.5,0.5,0.5}
\lstnewenvironment{code}{\lstset{language=Python}}{}

\definecolor{codegreen}{rgb}{0,0.6,0}
\definecolor{codepurple}{rgb}{0.58,0,0.82}
\definecolor{keywordsColor}{rgb}{0.000000, 0.000000, 0.635294}
\definecolor{backcolour}{rgb}{255,255,255}
\lstdefinestyle{mystyle}{
  backgroundcolor=\color{backcolour}, commentstyle=\color{codegreen},
  keywordstyle=\color{magenta},
  numberstyle=\tiny\color{codegray},
  stringstyle=\color{codepurple},
  basicstyle=\ttfamily\footnotesize,
  breakatwhitespace=false,         
  breaklines=true,                 
  captionpos=b,                    
  keepspaces=true,                 
  numbers=left,                    
  numbersep=5pt,                  
  showspaces=false,                
  showstringspaces=false,
  showtabs=false,                  
  tabsize=2,
  float=tp,
  floatplacement=tbp,
  xleftmargin=1.5em,
  belowskip=-8pt
}
\usepackage{xcolor}

\newcommand{\kamel}[1]{{\color{black}{#1}}\xspace}
\newcommand{\kamelv}[1]{{\color{black}{#1}}\xspace}
\newcommand{\NACCL}[1]{{\color{black}{#1}}\xspace}

\title{Can LLMs Patch Security Issues?}

\author{Kamel Alrashedy, Abdullah Aljasser,\\ \textbf{Pradyumna Tambwekar, Matthew Gombolay} \\
  Georgia Institute of Technology, GA, USA \\ 
  \texttt{\{kalrashedy3,aaljasser3,ptambwekar3\}}@gatech.edu \\
  \texttt{\{matthew.gombolay\}}@cc.gatech.edu}

\begin{document}
\maketitle
\begin{abstract}
Large Language Models (LLMs) have shown impressive proficiency in code generation. \kamelv{Unfortunately,} these models share a weakness with their human counterparts: \kamelv{producing code that inadvertently has security vulnerabilities.} These vulnerabilities \kamelv{could} allow unauthorized attackers to access sensitive data or systems, which is unacceptable for safety-critical applications. 
\kamelv{In this work,} we propose Feedback-Driven Security Patching (FDSP), wherein LLMs automatically refine vulnerable generated code.
Our approach leverages automatic static code analysis to empower the LLM to generate and implement potential solutions to address vulnerabilities. We address the research community’s needs for safe code generation by introducing a large-scale dataset, \textit{PythonSecurityEval}, covering the diversity of real-world applications, including databases, websites, and operating systems. We empirically validate that FDSP outperforms prior work that uses self-feedback from LLMs by up to 17.6\% through our procedure that injects targeted, external feedback.
Code and data are available at \url{https://github.com/Kamel773/LLM-code-refine}
\end{abstract}

\section{Introduction}
Although Large Language Models (LLMs), such as GPT-4 \citep{GPT3} and CodeLlama \citep{Code_llama}, are powerful tools for code generation, they are prone to generating vulnerable code \citep{Examining_Zero-Shot}. LLMs have shown high-competency for a wide \kamelv{variety of code generation tasks, such as for producing code from natural language} \citep{yu2018spider}, code translation \citep{lachaux2020unsupervised}, and code optimization \citep{shypula2023learning}.
Utilizing LLMs for code generation has been shown to increase developers' productivity with writing and explaining code, and fixing bugs \citep{wong2023natural}. To enhance code refinement with LLMs, recent work by \citet{self_debug} proposed a self-debugging technique, wherein LLMs generate code, which is then sent back to the same LLMs for evaluation and further refinement.

\begin{figure}
  \centering
 \fbox{%
   {\includegraphics[width=.96\linewidth]{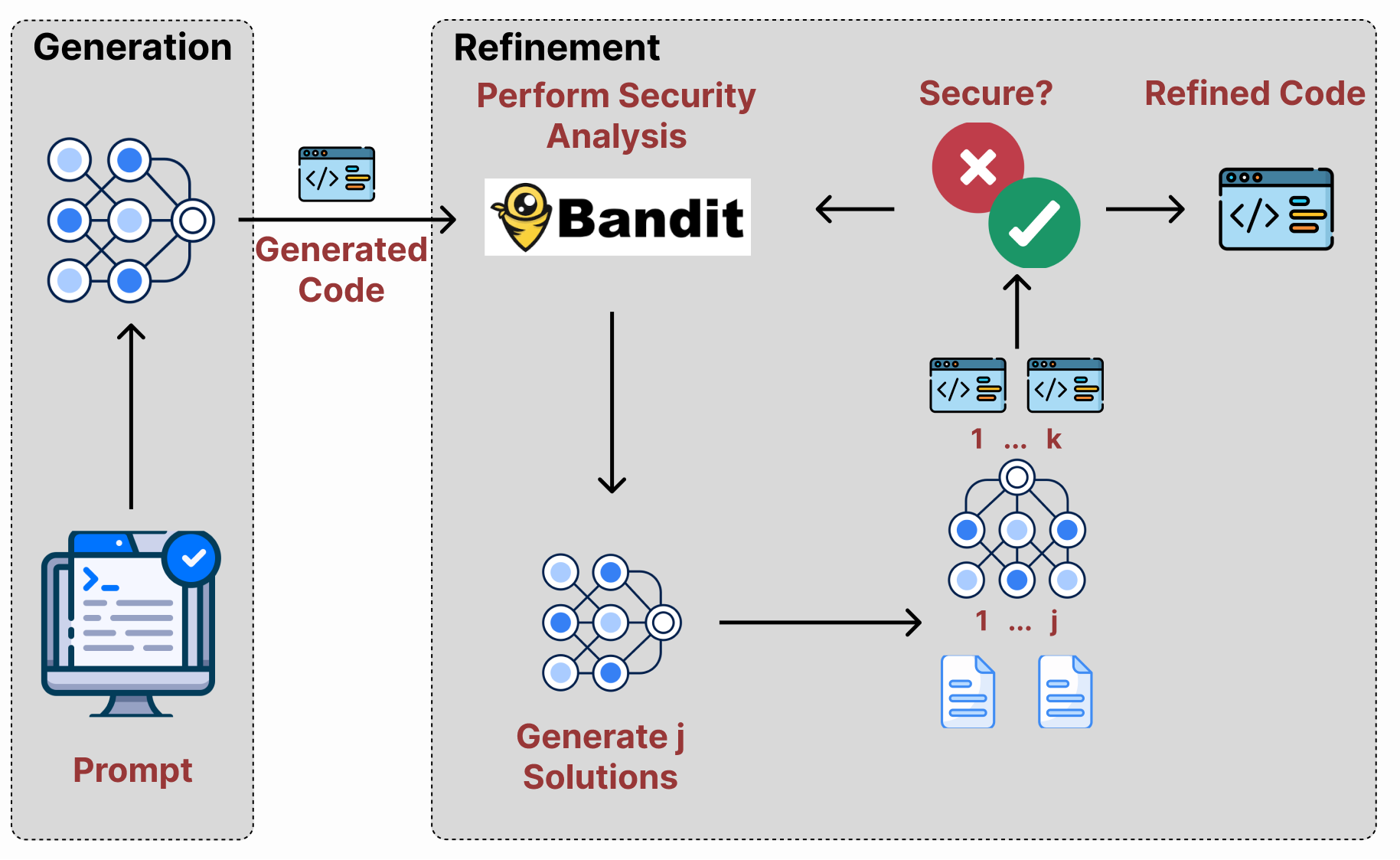}}
 }
  \caption{Overview of our approach: Initially, the LLMs generate code. This code is subsequently analyzed for security vulnerabilities using Bandit, a tool for static code analysis, to identify potential security issues (see Figure \ref{fig:Feedback_Bandit}). The identified potential security threats are then incorporated into the LLMs to generate possible solutions for resolving these security issues. Finally, each proposed solution is sent back to the LLMs for code refinement.}
  \label{fig:schematic}
\end{figure}

\kamelv{However, code generated  by LLMs could include security vulnerabilities. Vulnerabilities in code may allow unauthorized users to access sensitive data or systems. For example, attackers can manipulate SQL queries to gain access to a database, a technique known as SQL injection. 
Particularly, when the code interacts with external services and systems, LLMs might struggle to recognize and fix security issues in code due to their limited understanding of security vulnerabilities and lack of specific security knowledge \citep{Multi-lingual,Lightweight}.}

One potential approach to mitigate these security vulnerabilities is to train LLMs to recognize and patch security vulnerabilities. This method also has several significant challenges. Firstly, it requires a large human-labeled dataset that accurately distinguishes between vulnerabilities and non-vulnerabilities to train the LLMs effectively. Collecting such a human-labeled dataset is both costly and time-consuming. Additionally, there is a critical need for robust feedback mechanisms during the training process. LLMs require accurate feedback to learn from their mistakes, and this feedback must come from security experts.



\kamelv{In this paper, we address the key limitations of prior work by developing Feedback-Driven Security Patching (FDSP) and the \textit{PythonSecurityEval} benchmark. In FDSP,  LLMs generate potential solutions to ameliorate the security threats in the generated code. This process involves analyzing the generated code through static code analysis to identify potential  security threats and produce feedback. LLMs utilize this feedback to generate potential solutions, and then each potential solution solution, and its corresponding generated code, are returned to the LLMs to for further code refinement. Next, we curate an extensive benchmark from Stack Overflow, called \textit{PythonSecurityEval}, as existing security evaluation datasets are quite limited and insufficient to evaluate a model's ability to produce non-vulnerable code. Our dataset originates from real-world applications, providing diversity with prompts to generate code for a variety of applications, including databases (e.g., SQL, MySQL), URLs, operating systems, and websites (e.g., Flask). 
We consider these types of applications as primary sources of common vulnerabilities, including SQL injection, cross-site scripting (XSS), broken access control, and command injection.}

To summarize, our work presents three key contributions:
\begin{itemize}
    \item We propose Feedback-Driven Security Patching (FDSP), a technique that enhances LLMs capabilities to generate potential solutions for repairing security issues in the generated code by receiving feedback from (Bandit) static code analysis.

    \item  We present \textit{PythonSecurityEval}, a dataset consisting of $470$ natural language prompts designed to evaluate the ability of LLMs to generate secure code. 
        
    \item We demonstrate that FDSP outperforms prior works that use self-feedback by up to 17.6\% on the  \textit{PythonSecurityEval} dataset. We empirically evaluate the capabilities of the most advanced LLMs, including GPT-4, GPT-3.5, and CodeLlama, in identifying and refining insecure code. We utilize three benchmarks (including ours) and apply five baseline techniques for this evaluation.

\end{itemize}

\definecolor{lightblue}{rgb}{0.93, 0.95, 1.0}
\definecolor{lightorange}{rgb}{1.0, 0.9, 0.8}

\section{Related work}

\kamelv{We discuss two categories of previous work: LLMs for code generation and refinement of LLMs.}

\textbf{Language models for code:} \kamel{Code generation models have become a popular research area among Machine Learning (ML) and Software Engineering (SE) communities. 
The most common application of code generation models is the text-to-code generation task, wherein users prompt an LLM with natural language instructions to complete a coding task, and the LLM generates the corresponding code. Examples of the text-to-code generation include CodeLlama ~\citep{Code_llama} and CodeGeeX ~\citep{zheng2023codegeex}. All three achieve state-of-the-art performance on the Mostly Basic Programming Problems (MBPP) dataset ~\citep{austin2021program}. 
 The DocPrompting approach further demonstrates that prompting language models with code documentation improves code generation performance on models such as CodeT5, CodeX, and GPT-Neo on MBPP~\citep{zhou23docprompting}. 
 Beyond code generation, LLMs are also capable of code translation~\citep{roziere2020unsupervised}, code repair~\citep{allamanis2021self}, code documentation~\citep{nam2024using}, code testing~\citep{wang2024software} and defect prediction~\citep{alrashedy}.
 Our interest lies in exploring how these various capabilities demonstrated by LLMs can also be applied to address security issues in LLM-generated code.}

\textbf{Refinement of LLMs:}
\kamel{Recent studies have demonstrated that LLMs can refine their own output and adapt to feedback from external tools or human input. Self-Refine \citep{Self-refine} generates feedback and refines its output to improve the quality of the generated answers across 7 tasks using state-of-the-art models such as GPT-3.5 and GPT-4. 
Additionally, a similar technique called Self-Debugging \citep{self_debug} enables code generation models to debug initially generated code using feedback from the same LLM, unit test results, or compiler error messages. The feedback from the LLM explains the code line-by-line, which is then used to refine the generated code. This approach has shown improvement in three different code generation applications. An alternate approach, called Self-Repair ~\citep{olausson2023self}, seeks to produce feedback specifically focusing on why any faulty code snippet is incorrect. 
Another study \citep{gou2023critic} introduced CRITIC, which enables the model to engage with external tools such as a code interpreter, calculator, and search engine to receive feedback and improve the generated output.
In our work, we build on these self-refinement methods towards enabling large language models to fix security issues in generated code.
}


\kamelv{The feedback from the aforementioned methods come from multiple sources including human feedback, external tools, or environments.}
Human feedback is the most effective and accurate source of feedback
; however, it is also  costly and time-intensive \citep{Elgohary21NLEdit, Yuntao}. An alternative way to obtain feedback is from external tools such as compiler error messages for program repair \citep{yasunaga2020graph} and Pylint, a static code analyzer, for improving Python coding standards \citep{bafatakis2019python}. Additionally, previous studies have proposed techniques to obtain feedback from LLMs, including the LLM-Augmenter system \citep{peng2023check} and Recursive Reprompting and Revision framework \citep{yang2022re3}. Whereas the techniques described above utilize feedback receives from LLMs or external tools, our approach combines the strength of bother approaches by incorporating feedback from both external tools and LLMs, wherein the static code analysis provides feedback regarding the generated code, which LLMs utilize to generate potential solutions for addressing security threats in the code.

\begin{algorithm}[t]
\caption{FDSP algorithm}
\label{alg:one}
\begin{algorithmic}[1]
\Require Input $x$, LLMs $P_{LM}$, number of potential solutions $J$, number of iterations $K$
\Ensure Refine vulnerable code $y$ from the LLMs $P_{LM}(y_i|x)$
\State Initialize output $y_i$ from $P_{LM}(x)$
\\ // Generate potential solutions (Eqn. \ref{eq:generate_s})
\State $S \sim P_{LM}(y, \Re, j, p) $  
\\ //Iteration for each potential solution (Eqn. \ref{eq:iteration})
\For{$s \, \in \mathcal{S}$}
\For{$k \gets 1$ to $K$}                    
        \State $y_i \gets P_{LM}(y,s)$  
        \If{$\delta(y_i)$ is secure} \Comment{Stop condition}
    \State \textbf{Return} $y_i$ 
\EndIf
    \EndFor
\EndFor
\State \textbf{Return } $y$
\end{algorithmic}
\vspace{1em}
\end{algorithm}

\section{Our Approach}
\label{Sec:Our_Approach}
\color{black}Our approach, FDSP, both identifies and resolves vulnerabilities in code generated by an LLM. 
The principal component of FDSP is the use of static code analysis (Bandit) to generate solutions to potentially vulnerable code. 
We utilize a four-step approach: (i) code generation, (ii) code testing, (iii) solution generation and (iv) code refinement. 
The complete algorithm for FDSP is provided in Algorithm~\ref{alg:one}. \color{black}

\subsection{Code generation}

Given a natural language description of a Python function denoted as $x$, an LLM generates a Python program $y$ according to $P_{LM}(y|x)$. Next, the program $y$ is executed. If there is compiler error message, we send the program $y$ with $\{e_c\}$  to the LLMs to fix the error, as describe in Equation \ref{eq:fix}. The initial generated code can be described as follows:

\begin{equation}
y_i \sim P_{LM}(y_i|x) 
\label{eq:1}
\end{equation}

\begin{equation}
y^c \sim P_{LM}(y^c|x,y_i,e_c) 
\label{eq:fix}
\end{equation}

We consider a zero-shot setting because, in real-world use cases, users prompt LLMs to generate code directly without providing examples. We assumed that the LLM would generate secure code by default.

\subsection{Code testing}
Static code analysis tools are utilized by software engineers to evaluate the quality of the code and identify any potential vulnerabilities. We use Bandit,\footnote{https://github.com/PyCQA/bandit} a static code analysis tool designed to detect common vulnerabilities in Python functions. Bandit constructs the Abstract Syntax Tree\footnote{https://greentreesnakes.readthedocs.io/en/latest/} (AST) for a Python function and conducts analysis on the AST nodes. Subsequently, Bandit, denoted by $\delta$, generates a report $\Re$ about the code $y$ \kamelv{(see Figure \ref{fig:Feedback_Bandit}).}  Then, we pass the report $\Re$ from Bandit to the LLMs to generate potential solutions to fix the vulnerabilities. We can describe the Bandit report as follows:

\begin{equation}
\Re = \delta(y) 
\label{eq:2}
\end{equation}

\begin{lstlisting}[style=mystyle,escapechar=!,language=Python,label={example}, caption={An example of a prompt from PythonSecurityEval dataset where GPT-4 generates vulnerable code of the SQL injection type. The report from Bandit is shown in Figure \ref{fig:Feedback_Bandit}.}]
# The prompt: Write a Python function to return the total number of rows in SQLite.

import sqlite3
def get_total_rows(database_name, table_name):
    conn = sqlite3.connect(database_name)
    cursor = conn.cursor()
    cursor.execute("SELECT COUNT(*) FROM {}".format(table_name))
    result = cursor.fetchone()
    conn.close()
    return result[0]
\end{lstlisting}


\subsection{Generating potential solutions}
Static code analysis can detect vulnerabilities; however, it cannot fix them. 
The primary contribution of FDSP is to leverage static code analysis to enable LLMs to generate or suggest multiple solutions $S$ to address these vulnerabilities, as represented in Equation \ref{eq:generate_s}.




\begin{equation}
S \sim P_{LM}(y, \Re, j, p) 
\label{eq:generate_s}
\end{equation}

The variable $J$ represents the total number of generated potential solutions, and the $p$ is the instruction given to the LLMs. Figure \ref{fig:schematic} illustrates this process: the LLM generates code vulnerable to SQL injection, we then provide the LLM with report from Bandit, and subsequently, the LLM generates three different potential solutions: \textit{1) Use Parameterized Queries, 2) Manual Escape and Quote Table Name, and 3) Use an ORM (Object-Relational Mapping) Library} (see the full example in Figure \ref{fig:example_Generated_solution}). 



\begin{figure}[t]
\centering
\begin{tcolorbox}[title=An example of Bandit report., colback=gray!5, colframe=black, colbacktitle=gray!20, coltitle=black, sharp corners, width=\linewidth,]
 Issue: [B608:hardcoded\_sql\_expressions] Possible SQL injection vector through string-based query construction. \newline
\textit{Line 7:cursor.execute("SELECT COUNT(*) FROM {}".format(table\_name))}
\end{tcolorbox}
\caption{An example of the report generated by Bandit, a static code analysis tool, for the vulnerable code in Code Snippet \ref{example}.}
\label{fig:Feedback_Bandit}
\end{figure}

\subsection{Code refinement}
 We use the generated potential solutions  $S \sim  s_1  \oplus s_2  \ldots \oplus s_j$  from the previous step as feedback to fix the vulnerable code $y$. In the previous step the LLM generates $J$ unique solutions to fix the security issues. Then, each potential solution and vulnerable code is fed back into the LLM multiple times, denoted as $K$. The idea behind the $K$ iterations for each solution is to allow the LLM to generates as many fixes as possible for the vulnerable code. The refinement process is terminated when the Bandit detects no security issues or when the iteration reaches the maximum number of iterations, $K$, for all potential solutions $S$, as shown in Algorithm \ref{alg:one}.

\begin{equation}
y_{i+k} \sim \{\{P_{LM}(y_{i+k}|y_{i+k-1},s_n)\}_{n=1}^j\}_{i=1}^k 
\label{eq:iteration}
\end{equation}


\section{Experimental Settings}
In this section, we discuss the experimental setup used to evaluate the effectiveness of our proposed approach, FDSP.



\subsection{Benchmarks}
\label{Benchmarks}
Existing benchmarks, LLMSecEval and SecurityEval, are insufficient for large-scale evaluation due to their limited size and diversity (see Table \ref{tab:diver}). To address this limitation, we introduce \textit{PythonSecurityEval}, comprising 470 natural language prompts for diverse real-world applications, collected from Stack Overflow. We utilize \textit{PythonSecurityEval} to compare FDSP with existing strategies for fixing security issues.

\begin{itemize}
    \item \textbf{LLMSecEval:} This dataset contains natural language prompts to evaluate LLMs on generating secure source code \citep{llmseceval2023}. LLMSecEval is comprised of 150 total prompts (natural language descriptions of code), covering the majority of the top 25 Common Weakness Enumeration (CWE). 

    \item \textbf{SecurityEval:} This dataset, proposed by \citet{siddiq2022seceval}, is used to evaluate LLMs on their ability to generate secure Python programs.  SecurityEval comprises 121 natural language prompts covering 75 different types of vulnerabilities. Each prompt includes the header of a Python function along with comments describing the purpose of each function.

    \item \textbf{PythonSecurityEval (Ours):} We collected a new benchmark  from Stack Overflow to address the limitations of the existing datasets. Current datasets are limited in size and diversity; ergo are insufficient in evaluating the ability of LLMs to generate secure code adequately addressing security vulnerabilities.  \textit{PythonSecurityEval} includes natural language prompts intended to generate Python functions that cover diverse real-world applications. This dataset consisting of 470 prompts is three times larger than those used in LLMSecEval and SecurityEval.
    
\end{itemize}

\textit{PythonSecurityEval} is a diverse and extensive benchmark, covering the  majority of real-world applications that consider the primary sources of common vulnerabilities. For example, SQL injection occurs when Python code connects to, inserts into, and queries from a SQL database. There are several examples in our benchmark where the prompt involves writing Python code to insert a value into an SQL database. Another example is command injection, where the Python code interacts directly with the operating system, allowing attackers to gain unauthorized access to data and systems. We include examples that generate Python code with access to the operating system and diagnose how the LLMs generate the code without any issues or the ability to refine the code. Lastly, cross-site scripting (XSS) is a common type of security vulnerability that occurs in web applications. We include prompts that generate Python code for Flask, which is a Python web framework for creating websites (see Table \ref{tab:diver}).

\begin{table}[t]
\centering
\resizebox{0.47\textwidth}{!}{ 
\begin{tabular}{@{}l|ccc@{}}
\toprule
\multirow{2}{*}{\textbf{Domain}} & \textbf{PythonSecurityEval} & \textbf{LLMSecEval} & \textbf{SecurityEval} \\ 
                                 & \textbf{(Ours)} & \textbf{} & \textbf{} \\ 
\midrule
                         
                         Computation & 168 (35.7\%) & 44 (29.5\%) & 32 (26.4\%) \\
                          System & 313 (66.6\%) & 94 (63.1\%) & 68 (56.2\%) \\
                          Network & 147 (31.3\%) & 63 (42.3\%) & 29 (24.0\%)\\
                          Cryptography & 29 (6.2\%) & 8 (5.4\%) & 16 (13.2\%) \\
                         General & 414 (88.1\%) & 128 (85.9\%) & 118 (97.5\%) \\
                          Database & 114 (24.3\%) & 23 (15.4\%) & 6 (5.0\%) \\
                          Web Frameworks & 43 (9.1\%) & 46 (30.9\%) & 8 (6.6\%) \\   \cmidrule(lr){1-4}
                          Total & \textbf{470} & 150 & 121\\
\bottomrule
\end{tabular}
}
\caption{\NACCL{Comparison of domain diversity statistics for the distribution of functions in \textit{PythonSecurityEval} and prior benchmarks. The table shows the total number of functions identified per domain, along with their percentage distributions. Note that some functions are associated with multiple domains (see Table \ref{tab:python_modules}).}}
\label{tab:diver}
\end{table}

\subsection{Baselines}
\label{Baselines}
We consider four baseline refinement approaches, which are as follows:

\begin{enumerate}
    \renewcommand{\labelenumi}{\Roman{enumi}.}
    \item \textbf{Direct prompting:} This approach involves sending  generated code back to an LLM with the instruction: \textit{Does the provided function have a security issue? If yes, please fix the issue.} If LLMs detect any security issues in the code, they will fix the issue and generate secure code. 

    \item \textbf{Self-Debugging:} 
    The initial step in self-debugging is for LLMs to generate the code. Subsequently, the generated code is sent back to the same LLMs to generate feedback. Finally, both the generated code and the feedback are fed back to the LLM to correct any existing bugs. 


    \item \textbf{Bandit feedback:} We develop this baseline that utilizes Bandit to produce a report if there are any security issues in the code, as shown in Figure \ref{fig:Feedback_Bandit}. We use this report as feedback to enable the LLM to refine the vulnerable code. This strategy is similar to prior approached wherein external tools provide feedback to the LLM to refine its outputs \citep{gao2023pal, akyurek2023rl4f}. Bandit feedback does not provide a solution to fix the issue; it simply highlights the problematic line and type of issue.


\item \textbf{Verbalization:} \kamelv{
We verbalize the feedback from Bandit, via an LLM, to produce more understandable and actionable feedback to resolve security issues and defective code.
The verbalized feedback provides a detailed explanation in natural language of the specialized output from Bandit (see Figure \ref{fig:verbalization}).
This expanded explanation offers deeper insights into the security issues and may suggest solutions to address the vulnerabilities. 
}

\end{enumerate}



\subsection{Evaluation metrics}
\label{Evaluation}
\NACCL{This paper evaluates the ability of LLMs to generate vulnerable code and subsequently correct security issues identified in the code. We evaluate the accuracy of generated and refined vulnerable code by calculating the ratio of vulnerable code produced to the total amount of generated code. To verify whether the generated code contains vulnerabilities, we  utilize two evaluation metrics:} 

\begin{enumerate}
    \renewcommand{\labelenumi}{\Roman{enumi}}
     \setlength{\itemsep}{-2pt}
    \item \textbf{Bandit:} A static code analysis tool designed to identify vulnerabilities in Python code. \vspace{0.05cm}
    \item \textbf{CodeQL:} An open-source codebase utilized to discover the similarity of vulnerability patterns.\footnote{https://codeql.github.com/}

\end{enumerate}   

\subsection{Models}
\label{Models}
We aim to evaluate state-of-the-art LLMs for code generation, including GPT-4, GPT-3.5 \textit{``gpt-3.5-turbo-instruct"} \citep{GPT3} using OpenAI API, and CodeLlama \textit{``CodeLLama-Instruct-34B"}\citep{Code_llama} from Huggingface, to generate secure code. \NACCL{Additionally, we evaluate the ability of LLMs to refine insecure code using four baseline approaches (see \S \ref{Baselines}) and our proposed approach, FDSP.}





\subsection{Research Questions}
\NACCL{This paper explores four research questions, regarding the capacity of LLMs in detecting and refining vulnerable code.}

\begin{enumerate}
    \item[\textbf{RQ1.}] \textbf{What is the fundamental capability of LLMs in refining security vulnerabilities?} \NACCL{This question aims to determine how effectively LLMs can inherently correct insecure code and highlight their limitations without incorporating external feedback.}
    \item[\textbf{RQ2.}] \textbf{How does Bandit feedback affect the ability of LLMs to refine code vulnerabilities?} 
    \NACCL{This question examines how effectively the LLMs incorporate feedback provided by provided by Bandit, a static code analysis tool.}
    \item[\textbf{RQ3.}] \textbf{How does FDSP improve LLM performance in fixing code vulnerabilities?} 
    \NACCL{This question aims to assess how well the LLMs generate multiple potential solutions and iterate over each one to refine vulnerabilities.}
    \item[\textbf{RQ4.}] \textbf{How important are the multiple generated solutions and iterations of FDSP?} 
    \NACCL{We conduct ablation studies to isolate these factors by restricting FDSP to a single solution or iteration. This analysis reveals whether the diversity of generated solutions and iterative refinement contribute to FDSP effectiveness.}


\end{enumerate}

\begin{table*}[t]
\centering
\resizebox{\textwidth}{!}{
\begin{tabular}{@{}l|l|cc|cc|cc@{}}
\toprule
\textbf{Dataset} & \textbf{Models} & \multicolumn{2}{c}{\textbf{GPT 4}} & \multicolumn{2}{c}{\textbf{GPT 3.5}} & \multicolumn{2}{c}{\textbf{CodeLlama}} \\
 \cline{2-8}
& \textbf{Evaluation metrics} & \textbf{Bandit} & \textbf{CodeQL} & \textbf{Bandit} & \textbf{CodeQL} & \textbf{Bandit} & \textbf{CodeQL} \\
\midrule
\multirow{6}{*}{\textbf{LLMSecEval}} 
& Generated code & 38.2\% & 10.1\% & 34.2\% & 18.1\% & 28.6\% & 20.7\% \\
\cmidrule(lr){2-8}
& Direct prompting & 35.3\% $(\downarrow 2.6)$ & \textbf{4.7}\% $(\downarrow 5.4)$ & 28.0\% $(\downarrow 6.0)$ & 7.4\% $(\downarrow 10.7)$ & 24.0\% $(\downarrow 4.6)$ & 11.6\% $(\downarrow 9.1)$ \\
& Self-debugging & 24.0\% $(\downarrow 14.0)$ & 7.4\% $(\downarrow 2.7)$ & 28.0\% $(\downarrow 6.0)$ & 8.7\% $(\downarrow 9.4)$ & 24.6\% $(\downarrow 4.0)$ & 15.7\% $(\downarrow 5.0)$ \\
& Bandit feedback & 8.0\% $(\downarrow 30.0)$ & 5.4\% $(\downarrow 4.7)$ & 18.6\% $(\downarrow 15.3)$ & 8.7\% $(\downarrow 9.4)$ & 18.0\% $(\downarrow 10.6)$ & 13.2\% $(\downarrow 7.5)$ \\
& Verbalization & 7.3\% $(\downarrow 30.6)$ & 5.4\% $(\downarrow 4.7)$ & 18.0\% $(\downarrow 16.0)$ & \textbf{6.7\%} $(\downarrow 11.4)$ & 16.6\% $(\downarrow 12.0)$ & 10.7\% $(\downarrow 10.0)$ \\
& FDSP (Ours)& \textbf{6.0\%} $(\downarrow 32.0)$ & 6.7\% $(\downarrow 3.4)$ & \textbf{12.6\%} $(\downarrow 21.3)$ & 8.1\% $(\downarrow 10)$ & \textbf{14.6\%} $(\downarrow 14.0)$ & \textbf{9.1\%} $(\downarrow 11.6)$ \\
\midrule
\multirow{6}{*}{\textbf{SecurityEval}} 
& Generated code & 34.7\% & 12.4\% & 38.0\% & 9.1\% & 46.2\% & 32.2\% \\
\cmidrule(lr){2-8}
& Direct prompting & 21.4\% $(\downarrow 13.2)$ & \textbf{5.8\%} $(\downarrow 6.6)$ & 25.6\% $(\downarrow 12.4)$ & 8.3\% $(\downarrow 0.8)$ & 38.0\% $(\downarrow 8.2)$ & 14.1\% $(\downarrow 18.1)$ \\
& Self-debugging & 16.5\% $(\downarrow 18.1)$ & 9.1\% $(\downarrow 3.3)$ & 27.2\% $(\downarrow 10.7)$ & 9.1\% $(\downarrow 0.0)$ & 38.8\% $(\downarrow 7.4)$ & 17.4\% $(\downarrow 14.8)$ \\
& Bandit feedback & \textbf{4.1\%} $(\downarrow 30.5)$ & 6.6\% $(\downarrow 5.8)$ & 13.2\% $(\downarrow 24.7)$ & 5.8\% $(\downarrow 3.3)$ & 21.4\% $(\downarrow 24.7)$ & 13.4\% $(\downarrow 18.8)$ \\
& Verbalization & 4.9\% $(\downarrow 29.7)$ & 6.6\% $(\downarrow 5.8)$ & 13.2\% $(\downarrow 24.7)$ & 5.8\% $(\downarrow 3.3)$ & 17.3\% $(\downarrow 28.92)$ & 13.4\% $(\downarrow 18.8)$ \\
& FDSP (Ours) & \textbf{4.1\%} $(\downarrow 30.5)$ & 8.3\% $(\downarrow 4.1)$ & \textbf{5.7\%} $(\downarrow 32.2)$ & \textbf{2.5\%} $(\downarrow 6.6)$ & \textbf{8.2\%} $(\downarrow 38.0)$ & \textbf{12.1} $(\downarrow 20.1)$ \\
\midrule
\multirow{6}{*}{\textbf{PythonSecurityEval}} 
& Generated code & 40.2\% & 17.9\% & 48.5\% & 13.2\% & 42.3\% & 13.2\% \\
\cmidrule(lr){2-8}
& Direct prompting & 25.1\% $(\downarrow 15.1)$ & 9.6\% $(\downarrow 8.3)$ & 42.5\% $(\downarrow 5.9)$ & 8.5\% $(\downarrow 7.2)$ & 31.0\% $(\downarrow 11.3)$ & 6.6\% $(\downarrow 6.6)$ \\
& Self-debugging & 24.8\% $(\downarrow 15.3)$ & 8.7\% $(\downarrow 9.2)$ & 43.4\% $(\downarrow 5.1)$ & 8.9\% $(\downarrow 7.0)$ & 33.1\% $(\downarrow 9.2)$ & 7.9\% $(\downarrow 5.3)$ \\
& Bandit feedback & 9.3\% $(\downarrow 30.8)$ &  9.1\% $(\downarrow 8.8)$ & 26.3\% $(\downarrow 22.1)$ & 6.4\% $(\downarrow 11.5)$ & 20.0\% $(\downarrow 22.3)$ & 6.2\% $(\downarrow 7.0)$ \\
& Verbalization & 8.7\% $(\downarrow 31.4)$ & 8.5\% $(\downarrow 9.4)$ & 23.6\% $(\downarrow 24.8)$ & 7.4\% $(\downarrow 10.2)$ & 19.5\% $(\downarrow 22.8)$ & 6.0\% $(\downarrow 7.2)$ \\
& FDSP (Ours) & \textbf{7.4\%} $(\downarrow 32.7)$ & \textbf{7.7\%} $(\downarrow 10.2)$ & \textbf{15.7\%} $(\downarrow 32.7)$ & \textbf{5.7\%} $(\downarrow 11.7)$ & \textbf{8.7\%} $(\downarrow 33.6)$ & \textbf{5.7\%} $(\downarrow 7.5)$ \\
\bottomrule
\end{tabular}
}
\caption{The table presents our results from Bandit and CodeQL evaluations metrics across various datasets and approaches. The percentage of vulnerable code is reported for both the generation and refinement phases. The (Generated code) row indicates the percentage of vulnerable code out of the entire dataset. We report the percentage of remaining vulnerable code for each refinement approach with the number in parentheses representing the percentage of vulnerable code that was fixed relative to the (Generated code).}
\label{tab:main_results}
\end{table*}

\section{Experimental Results}
In this section, we present the empirical evaluations pertaining to each research question and report the results in in Table \ref{tab:main_results}.

\textbf{LLMs are somewhat effective at refining vulnerable code on their own (RQ1).} 
Direct prompting and self-debugging are refinement methods where LLMs refine their generated code without feedback from external tools. Across the three datasets, Bandit detects between 28\% and 46\% of the generated code as vulnerable, while CodeQL identifies between 9.1\% and 17\%. In Bandit's evaluation, direct prompting and self-debugging result in modest vulnerability reductions, with improvements of less than 10\% for GPT-3.5 and CodeLlama, and around 15\% for GPT-4. This indicates that LLMs can intrinsically generate feedback to refine their vulnerable code, though the improvement is limited. CodeQL detects fewer vulnerabilities than Bandit, but results indicate that direct prompting and self-debugging fix approximately 50\% of the vulnerabilities.


\kamelv{\textbf{Bandit-based feedback is beneficial towards correcting security vulnerabilities in generated code (RQ2).}} Integrating Bandit feedback into LLMs enhances their ability to address security vulnerabilities, as evidenced by notable improvements in Bandit evaluations and modest gains in CodeQL evaluations. In contrast, approaches that exclude Bandit’s feedback are less effective. While simple strategies like direct prompting and self-debugging can address  basic security issues, they are generally insufficient for more complex vulnerabilities. As shown in Table \ref{tab:main_results}, methods utilizing Bandit feedback consistently outperform simpler techniques, improving accuracy across all models and datasets. Specifically, LLMs incorporating Bandit feedback provides approximately a 30\% improvement for GPT-4 and up to a 24\% improvement for GPT-3.5 and CodeLlama based on Bandit evaluations. Additionally, verbalizing Bandit's feedback yields a slight increase of 1\% to 2\% in both evaluation metrics.

\kamelv{\textbf{FDSP shows consistent improvement over the baseline (RQ3).}} FDSP boosts the LLM ability to generate potential solutions based on feedback provided by Bandit. Our FDSP approach enhances the performance of GPT-3.5 and CodaLlama, exceeding the results achieved by either directly incorporating Bandit's feedback or verbalizing it. For \textit{PythonSecurityEval}, FDSP shows consistent improvement over the verbalization approach, with improvements for GPT-4 (from 8.7\% to 7.4\%), GPT-3.5 (from 23.6\% to 15.7\%), and CodeLlama (from 21\% to 13.6\%) for Bandit evaluation. These results support that LLMs can propose potential solutions and provide useful feedback to fix security issues when they are supplied with feedback from static code analysis, and outperforming self-refinement or merely passing the feedback from static code analysis directly.

\begin{figure}[t]
  \centering
 {\includegraphics[width=.99\linewidth]{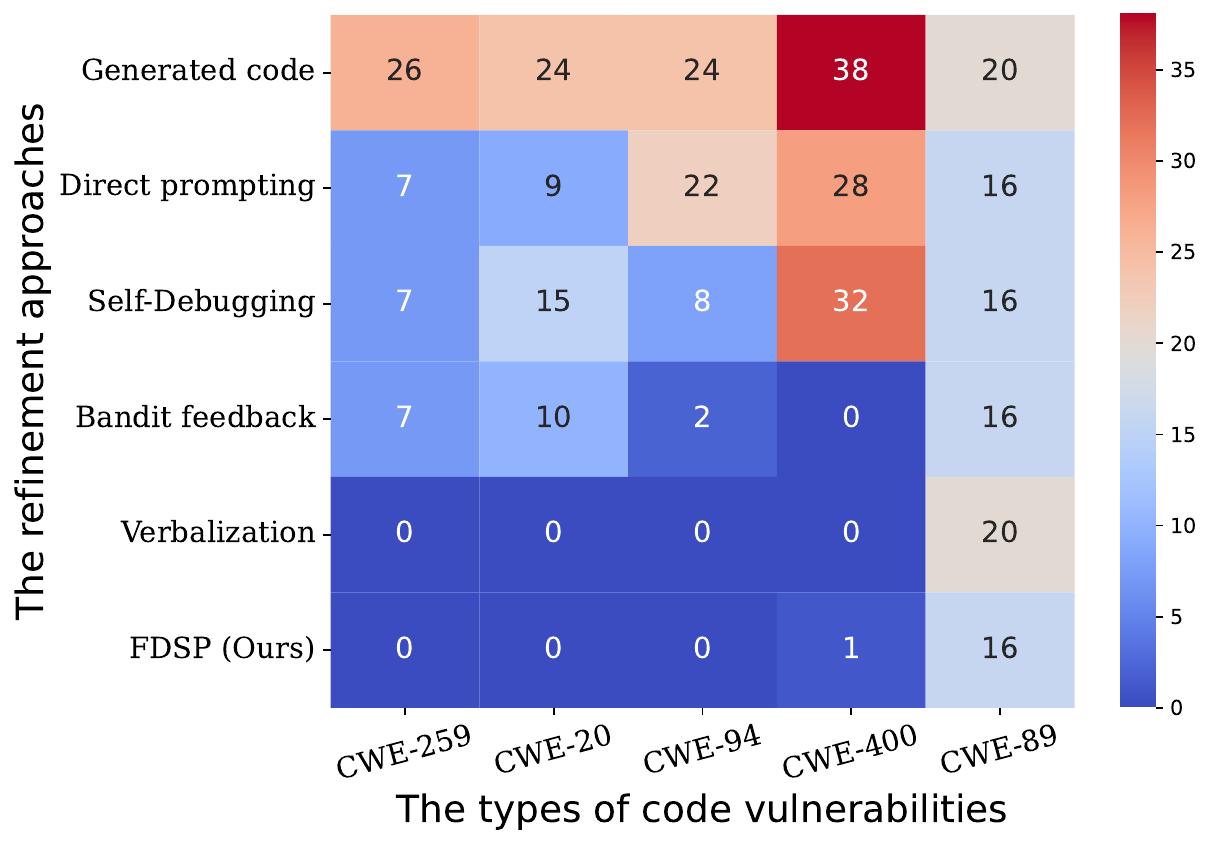}}
  \caption{\NACCL{The total count of the five most frequent security issues across five refinement approaches for CodeLlama in the \textit{PythonSecurityEval} dataset.}}
  \label{fig:Compare_approach_CodeLlam}
\end{figure}

We evaluate the effectiveness of each method in addressing the most common security issues in CodeLlama (see Figure \ref{fig:Compare_approach_CodeLlam}). These results suggest that neither self-refinement nor directly passing the feedback from Bandit proves useful for CodeLlama; however, verbalization and FDSP performs well for CodeLlama.

\begin{table}[t]
\centering
\small 
\begin{tabular}{@{}l|cc@{}}
\toprule
\multirow{2}{*}{\textbf{Ablation experiments}}
& \multicolumn{2}{c}{\textbf{Evaluation metrics}}  \\
\cmidrule(lr){2-3}
& \textbf{Bandit} & \textbf{CodeQL} \\
\midrule
Generated code & 40.2\% & 17.9\% \\
\cdashline{1-3}[2pt/2pt]
\rule{0pt}{10pt}FDSP with single solution & 10.0\% (\textcolor{red}{+2.6\%}) & 8.7\% (\textcolor{red}{+1.0\%})\\
FDSP with single iteration & 9.5\% (\textcolor{red}{+2.1\%})& 7.9\% (\textcolor{red}{+0.2\%})\\
\midrule
FDSP & \textbf{7.4\%} & \textbf{7.7\%} \\
\bottomrule
\end{tabular}
\caption{\NACCL{Performance comparison of FDSP and its ablated variants on the \textit{PythonSecurityEval} dataset using GPT-4. The table demonstrates FDSP's effectiveness across multiple solutions and iterations.}}
\label{tab:ablation}
\end{table}

\textbf{Ablation study (RQ4).} \NACCL{The generation of multiple solutions and repeated iterations play a critical role in FDSP (see \S \ref{Sec:Our_Approach}). To quantify the impact of these two factors, we evaluate FDSP with two ablation studies: (i) FDSP with a single solution, wherein the LLM generates only one solution instead of multiple, and (ii) FDSP with a single iteration, wherein the LLM attempts to address the vulnerable code with a single generated solutions rather than making multiple attempts. We conducted these ablation studies using GPT-4 on the \textit{PythonSecurityEval} dataset (see Table \ref{tab:ablation}). The percentage of unsolved vulnerabilities increased from 7.4\% to 9.5\% for single iteration and 10.0\% for single solution. 
These results demonstrate that while multiple iterations offer some improvement, generating multiple solutions plays a more significant role in improving the generation of secure code.}

\begin{figure}[t]
  \centering
 {\includegraphics[width=.99\linewidth]{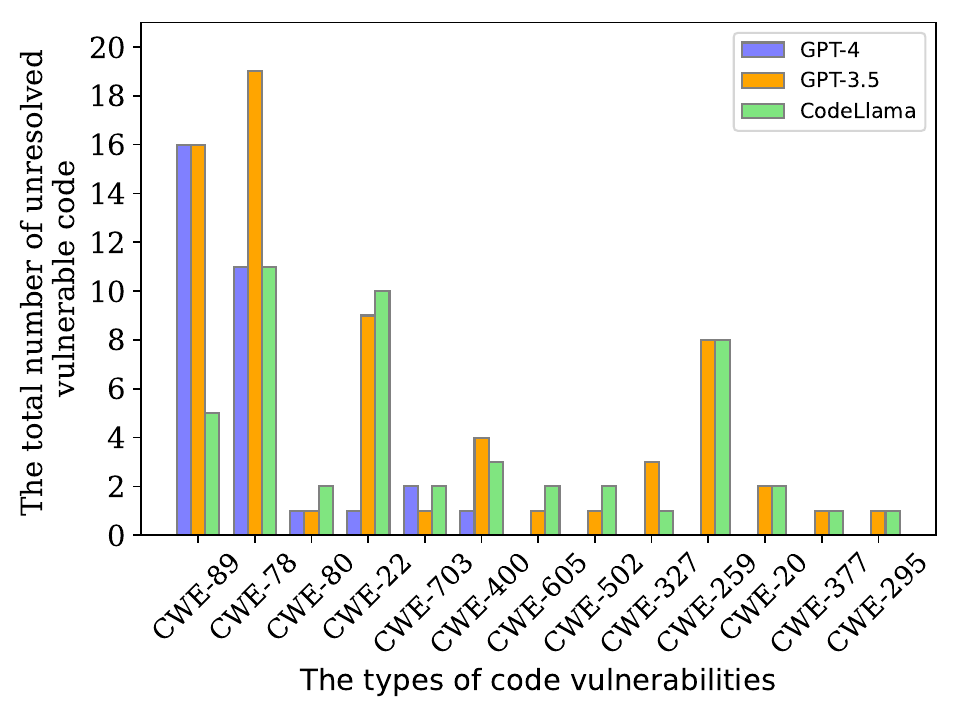}}
  \caption{\NACCL{Comparison of the total number of unresolved vulnerable code instances identified by three LLMs on the PythonSecurityEval dataset.}}
  \label{fig:unresolved_security_issues}
\end{figure}

\textbf{Qualitative analysis.} \NACCL{We perform a qualitative analysis of the solutions generated by FDSP and its iterations for GPT-4. In particular, we manually evaluate 30 randomly selected examples of vulnerable code from \textit{PythonSecurityEval}, comprising 23 fixed and 7 unfixed cases. Our findings show that the solutions generated by FDSP provide at least one actionable fix to address security issues, with 74\% of the solutions offering at least two actionable recommendations. Although FDSP consistently generates correct solutions, there are instances where the LLMs fail to refine the vulnerable code. Only 26\% of the generated solutions include one general security measure, such as error handling (e.g., exceptions) or input validation

In the seven cases where the vulnerabilities were not fixed, FDSP still produced valid potential solutions, but the LLMs did not incorporate the feedback to refine the code. Three of these failures involved SQL injection vulnerabilities, where FDSP produced valid potential solutions, but the LLMs failed to incorporate the feedback and refine the code (see Table \ref{tab:potential_solutions}). The other four failures involved high-risk library calls (e.g., subprocess, paramiko), which pose significant security risks if not used properly, potentially leading to shell injection vulnerabilities. Nevertheless, in the fix cases, FDSP generated accurate and useful solutions, with the vast majority of vulnerabilities being resolved in the first iteration.}

\begin{figure}[t]
  \centering
 {\includegraphics[width=.99\linewidth]{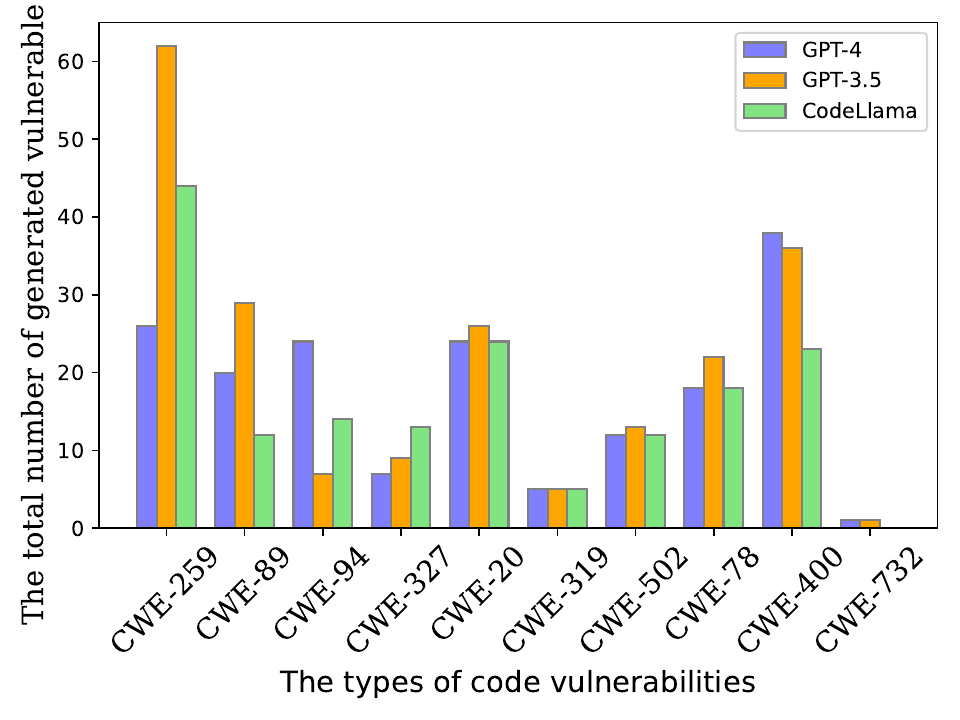}}
  \caption{\NACCL{The total count of the most common security issues in the code generated for the \textit{PythonSecurityEval} dataset (Top 10).}}
  \label{fig:Total_security_issues}
\end{figure}

\kamelv{\textbf{What are the most frequent and unresolved coding vulnerabilities produced by LLMs.}} We analyze the most common vulnerabilities in generated code, as well as those that remain unresolved, for the \textit{PythonSecurityEval} dataset. Figure \ref{fig:Total_security_issues} illustrates the most common types of code vulnerabilities generated by three LLMs, with the top two being CWE-259 (use of Hard-coded Password) and CWE-400 (uncontrolled resource consumption). However, the LLMs are able to fix most of these types of vulnerabilities (see Figure \ref{fig:unresolved_security_issues}). We visualize the most frequent unresolved security issues in Figure \ref{fig:unresolved_security_issues}, where the top two are related to injection: CWE-78 (OS Command Injection) and CWE-89 (SQL Injection), with percentage of 61.1\% and 80.0\% respectively for GPT-4. Additionally, these injection vulnerabilities are also among the most frequent vulnerabilities generated by LLMs.

\section{Conclusion}
As LLMs become capable of generating code, we assess the frequency with which they produce vulnerable code. We conduct an empirical evaluation of LLMs using existing datasets and approaches, and discuss their limitations. We propose a novel approach and introduce a new dataset to address the current limitations of both datasets and approaches. Our approach, FDSP, wherein the LLMs generate potential solutions to fix their own generated vulnerable code by leveraging feedback from static code analysis. Our results demonstrate that the FDSP approach outperforms the baselines across all three benchmarks and three models, achieving an improvement of about 17.6\% in the \textit{PythonSecurityEval} dataset for GPT-4.

\section{Limitations}
One of the limitations of our study is that our evaluation may not identify all security issues in the code. Detecting vulnerabilities in code is an active area of research, as recent studies have shown that some vulnerabilities cannot yet be detected by either static code analysis or deep learning-based vulnerability detection methods \citep{chakraborty2021deep}. Additionally, we evaluate the capabilities of LLMs in generating and refining vulnerable code at the function level; however, in real-life scenarios, vulnerabilities often occur when code interacts with other code or services. Lastly, our approach to fixing security issues involves making changes to the code, which has the potential to inadvertently render the program incorrect. It is very challenging to ensure the correctness of the code or to write unit tests, as all the generated code interacts with other services or systems such as databases or operating systems.


\section*{Acknowledgments}
We would like to thank Aman Madaan for his contributions to this paper. Aman contributed in the following ways: (1) developing Algorithm 1, (2) providing feedback on the writing of this paper, and (3) offering helpful discussions regarding the baseline and related works.

\bibliography{acl_latex}

\onecolumn
\newpage

\appendix

\section{Appendix}
\label{sec:appendix}

\begin{figure}[!htb]
\centering
\begin{tcolorbox}[title=An example of verbalization., colback=gray!5, colframe=black, colbacktitle=gray!20, coltitle=black, sharp corners, width=.99\linewidth,]
The feedback from the static code analysis regarding the issue tagged as "B608: hardcoded sql expressions" indicates a security concern with the way SQL queries are being constructed in the code. Specifically, the issue points out that the code might be vulnerable to SQL injection attacks due to the use of string-based query construction. This vulnerability arises when input data is concatenated directly into SQL queries without proper sanitization or parameterization, allowing an attacker to manipulate the queries to execute arbitrary SQL commands.

The problematic code is found at line 7, where a cursor's execute method is called to perform a SQL query. The query attempts to count the rows in a table by dynamically inserting the table name into the SQL statement. However, the method of constructing this query by formatting a string is unsafe if the table name comes from an untrusted source or is not properly validated, as it can be exploited to inject malicious SQL.

To mitigate this risk, it's recommended to use parameterized queries or prepared statements provided by the database interface, which ensure that input data is treated as parameters rather than part of the SQL command. This approach helps prevent SQL injection by separating the data from the code that queries the database.
\end{tcolorbox}
\caption{An example of how GPT-4 verbalizes the feedback from the Bandit shown in the Figure \ref{fig:Feedback_Bandit}.}
\label{fig:verbalization}
\end{figure}

\begin{figure}[!htb]
\centering
\begin{tcolorbox}[title=An example of generated solution., colback=gray!5, colframe=black, colbacktitle=gray!20, coltitle=black, sharp corners, width=.99\linewidth,]
\textbf{1) Use Parameterized Queries:} Parameterized queries ensure that user input is treated as a literal value rather than
executable code. Most database libraries provide a way to create these queries, also known as prepared statements.

\textbf{2) Manual Escape and Quote Table Names:} Since parameterized queries do not support table or column names, you
can manually ensure that table names are valid, using a whitelist approach where only approved table names are used.

\textbf{3) Use an ORM (Object-Relational Mapping) Library:} ORMs provide an abstraction over SQL by allowing you to interact as SQLAlchemy for Python handle escaping and quoting internally in a secure manner
\end{tcolorbox}
\caption{An example of a solution generated for the security issues in Figure \ref{fig:Feedback_Bandit}.}
\label{fig:example_Generated_solution}
\end{figure}

\begin{table*}[h]
\centering
\small
\resizebox{\textwidth}{!}{ 
\begin{tabular}{@{}lp{0.65\textwidth}@{}}
\toprule
\textbf{Domain} & \textbf{Library} \\ 
\midrule
\textbf{Computation} & os, pandas, numpy, sklearn, scipy, math, nltk, statistics, cv2, statsmodels, tensorflow, sympy, textblob, skimage \\ \cmidrule(lr){1-2}
\textbf{System} & os, json, csv, shutil, glob, subprocess, pathlib, io, zipfile, sys, logging, pickle, struct, psutil \\ \cmidrule(lr){1-2}
\textbf{Network} & requests, urllib, bs4, socket, django, flask, ipaddress, smtplib, http, flask\_mail, cgi, ssl, email, mechanize, url \\ \cmidrule(lr){1-2}
\textbf{Cryptography} & hashlib, base64, binascii, codecs, rsa, cryptography, hmac, blake3, secrets, Crypto \\ \cmidrule(lr){1-2}
\textbf{General} & random, re, collections, itertools, string, operator, heapq, ast, functools, regex, bisect, inspect, unicodedata \\ \cmidrule(lr){1-2}
\textbf{Database} & sqlite3, mysql, psycopg2, sqlalchemy, pymongo, sql \\ \cmidrule(lr){1-2}
\textbf{Web Frameworks} & Django, Flask, FastAPI, Tornado, Pyramid, Bottle \\ 
\bottomrule
\end{tabular}
}
\caption{We determined the type of domain for the function presented in Table \ref{tab:diver} by identifying calls to domain-specific libraries \citep{zhuo2024bigcodebench}}
\label{tab:python_modules}
\end{table*}

\begin{table}[h]
\centering
\small 
\begin{tabular}{@{}c|l@{}}
\toprule
\multirow{1}{*}{\textbf{CWE ID}} & \multicolumn{1}{c}{\textbf{Description}} \\
\cmidrule(lr){1-2}
CWE-20 & Improper Input Validation \\
CWE-22 & Improper Limitation of a Pathname to a Restricted Directory ('Path Traversal') \\
CWE-78 & Improper Neutralization of Special Elements used in an OS Command ('OS Command Injection') \\
CWE-79 & Improper Neutralization of Input During Web Page Generation ('Cross-site Scripting') \\
CWE-89 & Improper Neutralization of Special Elements used in an SQL Command ('SQL Injection') \\
CWE-94 & Improper Control of Generation of Code ('Code Injection') \\
CWE-119 & Improper Restriction of Operations within the Bounds of a Memory Buffer \\
CWE-200 & Exposure of Sensitive Information to an Unauthorized Actor \\
CWE-284 & Improper Access Control \\
CWE-287 & Improper Authentication \\
CWE-306 & Missing Authentication for Critical Function \\
CWE-352 & Cross-Site Request Forgery (CSRF) \\
CWE-400 & Uncontrolled Resource Consumption \\
CWE-502 & Deserialization of Untrusted Data \\
CWE-611 & Improper Restriction of XML External Entity Reference (XXE) \\
CWE-703 & Improper Handling of Exceptional Conditions \\
CWE-798 & Use of Hard-coded Credentials \\
CWE-120 & Buffer Copy without Checking Size of Input ('Classic Buffer Overflow') \\
CWE-125 & Out-of-bounds Read \\
CWE-190 & Integer Overflow or Wraparound \\
CWE-295 & Improper Certificate Validation \\
CWE-416 & Use After Free \\
CWE-434 & Unrestricted Upload of File with Dangerous Type \\
CWE-476 & NULL Pointer Dereference \\
CWE-732 & Incorrect Permission Assignment for Critical Resource \\
CWE-787 & Out-of-bounds Write \\
CWE-862 & Missing Authorization \\
CWE-918 & Server-Side Request Forgery (SSRF) \\
CWE-1021 & Improper Restriction of Rendered UI Layers or Frames \\
CWE-1295 & Debug Features Enabled in Production \\
\bottomrule
\end{tabular}
\caption{Common Weakness Enumeration (CWE) IDs and descriptions.}
\label{tab:cwe_list}
\end{table}

\begin{table*}[h]
\centering
\small
\resizebox{\textwidth}{!}{ 
\begin{tabular}{@{}lp{0.65\textwidth}@{}} 
\toprule
\textbf{Examples} & \textbf{Generated potential solutions} \\ 
\midrule
\textbf{Example 1} & \textbf{1) Use Parameterized Queries:} The most effective way to prevent SQL injection is to use parameterized queries, also known as prepared statements. This strategy doesn't form SQL commands by using string operations. Instead, it designates placeholders for each parameter, and the database engine will fill these placeholders securely. However, SQLite doesn't allow placeholders for table names, so it might not be possible to bind table names to placeholders in this case.

\textbf{2) Input Validation:}
Another strategy is to validate the user inputs or in this case, the table\_name variable. We can ensure that it only contains valid table name characters and does not contain dangerous SQL syntax. This can be done using a whitelist of allowed characters or a regular expression (regex) to match valid table names.

\textbf{3) Devise a Whitelisting Mechanism:}
This would involve creating a list of tables that are allowed to be accessed via the "total\_rows" function. By checking if the table name provided to the function is present in the whitelisted table names, the code can ensure that only legitimate queries are executed, proving effective against SQL injection.  \\ 
\midrule
\textbf{Example 2} & 
\textbf{1)Parameterized Query:} The main solution for preventing SQL injection is to use parameterized queries. This entails using placeholders for data in your SQL statements and then providing the data to the SQL interpreter separately so it's correctly and safely quoted. Unfortunately, SQLite in Python doesn't directly support parameterization of table or column names. So, the programmer needs to ensure they sanitize their table or column name inputs.

\textbf{2) Use of Sanitizing function:} To handle not directly supported parameterization, we can also create a sanitizing function to ensure the validity of the table name provided to prevent SQL injection though still possible to think about the limitation of sanitizing process.

\textbf{3) Use of Database API:} Instead of allowing the program user to provide arbitrary table or column names, consider using a Database API to generate these names. This will provide an interface for the user to view and interact with the data, but not change the structure of the SQL queries themselves. This significantly reduces the risk of SQL injection attacks because the user may only affect the value in a field, not the structure of the SQL queries since the table names are not user inputs.

\\ 
\midrule
\textbf{Example 3} &   \textbf{1) Use Parameterized Queries:} This involves using placeholders instead of directly inserting values into the statements. Parameters are used to help differentiate between the SQL code and the data passed through. When executed, the SQL statement compiles together with the parameters as one command which prevents attackers from interfering with the structure of SQL queries.

\textbf{2) Use Object-Relational Mapping (ORM) Libraries:} Technologies such as SQLAlchemy, Django ORM, can be used to handle operations with databases. These libraries usually come with secure features that protect from SQL injection attacks.

\textbf{3) Use Database Views:} Business logic could be implemented in the database side to decrease the complexity of SQL built in the application, and to encapsulate the database structure. Views could be used to give simplified perspective of the database and to restrict access to sensitive parts of the database schema. Reducing database control and visibility minimizes the potential damage of SQL Injection attacks.
\\
\bottomrule
\end{tabular}
}
\caption{Examples of corrected potential solutions generated by FDSP for SQL injection vulnerabilities, where GPT-4 was unable to refine the vulnerable code.}
\label{tab:potential_solutions}
\end{table*}

\end{document}